\documentclass[preprint,12pt]{elsarticle}

\usepackage{amsmath}
\usepackage{amssymb}
\usepackage{stmaryrd}
\usepackage{stackrel}
\usepackage{graphicx}
\usepackage{multicol}
\usepackage{url}
\usepackage{hyperref}
\hypersetup{colorlinks}
\usepackage{subfig}

\begin{document}

\begin{frontmatter}


\title{Real particle geodesics and thermodynamics of a black hole in Regular Schwarzschild-Anti de Sitter space-time}


\author{\textit{A. Mahmoodzadeh}\url{A. Mahmoodzadeh@iau.ac.ir},\hspace{5cm}\textit{B. Malekolkalami}\url{B.Malakolkalami@uok.ac.ir},\hspace{5cm} \textit{K. Ghaderi}\url{K. Ghaderi@iau-Marivan.ac.ir} }

\address{Department of Physics, Boukan Branch, \textbf{Islamic Azad University}, Boukan, Iran}
\address{Faculty of Science, \textbf{University of Kurdistan}, Sanandaj, P. O. Box 416, Iran}
\address{Department of Physics, Marivan Branch, \textbf{Islamic Azad University}, Marivan, Iran}

\begin{abstract}
  Regular Schwarzschild Anti-de-Sitter (RSch-AdS) spacetime, which is a maximally symmetric and Lorentzian manifold of negative curvature, is a solution of Einstein's field equations with a negative cosmological constant and a matter source with a Gaussian distribution. Studying of black hole (BH) in such a spacetime, known as RSch-AdS black hole, can be instructive from various aspects and lead to a better understanding of them. In this regard, geodesics of real particles around RSch-AdS black holes and RSch-AdS thermodynamics of BH are two important issues that we are going to evaluate in this research. For this purpose, we first obtain the geodesics of real particles based on the changes in the values of effective parameters such as mass distribution, angular momentum and cosmological constant numerically according to boundary conditions and interpret their behavior. Second, using the laws of thermodynamics, some aspects of BH, defined in this spacetime, such as temperature, entropy, heat capacity and Gibbs free energy are studied and discussed. 
\end{abstract}


\end{frontmatter}

\section{Introduction}
\label{sec:1}
   Since Hawking's announcement in 1975 about black hole evaporation\cite{key-1}, overwhelming and continuous efforts have been made in this matter, and there is still a long way to go before solving all the existing challenges and reaching the final results to have a clear understanding of black hole behaviors. To remind the readers, some of these challenges are listed here. 1.- What happens in the final stage of evaporation and the possible loss of information encoded in quantum state of matter?. 2.- Hawking radiation is negligible for astrophysical black holes because  their temperature can reach at most some tens of $nK$, so far below $T_{CMB}\sim 2.7$ and data do not provide acceptable results. 3.- Although there are many evidence that prove the existence of black holes, but identification of them with mass less than three times the mass of solar is unclear and relevance of these objects is connected with the possibility of observing the Hawking radiation. 4.- Divergent behavior of Hawking temperature for black holes created by high energy particle collision can be described by Schwarzschild geometry which has a curvature singularity at the origin, but why in practice we don't expect such an event in the vicinity of origin?. For a relatively complete overview on these topics we refer you to \cite{key-2} and references therein.    
   
   We recall that a static and spherically symmetric black hole described by a Schwarzschild geometry, due to losing mass during its evaporation, shrinks to its origin where there is a singularity, and therefore, there would be a divergent behavior of Hawking temperature. However, strong quantum gravitation fluctuation of the spacetime manifold in the vicinity of the origin prevents from taking place the divergence and a Plank phase of the evaporation will happen in this period which Quantum Gravity and Quantum Field Theory, a conjectured relationship between AdS sapcetime on one side that are used in quantum gravity formulated in terms of string theory or M-theory and on the other side Conformal Field Theory including theories similar to the Yang-Mills theories that describe elementary particles, said AdS/CFT correspondence, must be invoked\cite{key-3,key-4,key-5,key-6,key-7}. The problem here is that the aforementioned theories still cover few cases and Loop Quantum Gravity suffer from the absence of a clear semiclassical limit. Beyond the rough semiclassical approximation of QFT in curved space, the more significant approach is noncommutative geometry which describes black hole based on microscopic statistical nature of gravity. On the other hand studying the thermodynamics of black hole,in a spacetime known as Sch-AdS spacetime, based on noncommutative geometry framework is an outstanding description of quantum gravitational spacetime \cite{key-8}.  
   
      Therefore, in order to have a general view of the AdS spacetime, this section ends with a brief definition. The AdS spacetime is a differential (Lorentzian) manifold of dimension 4, one of the simplest, most maximally spherical symmetric and a static solution of Einstein's equations described by $R_{\mu\nu}-\frac{1}{2} Rg_{\mu\nu}+\Lambda g_{\mu\nu}=\frac{8\pi G}{c^4}T_{\mu\nu}$, where $R_{\mu\nu}$ is the Ricci curvature tensor and can be determined by mass-energy contents, $R$ is Ricci scalar curvature, $g_{\mu\nu}$ is the metric tensor of spacetime, $\Lambda$ is negative cosmological constant and $T_{\mu\nu}$ the momentum-energy \cite{key-3,key-9,key-10}. 

The paper is generalized as bellow: In Section \ref{sec:2}, a mathematical brief of noncommutative coordinates resulting in RSch-AdS spacetime will be mentioned. In Section \ref{sec:3}, geodesics plots for this spacetime are calculated numerically based on time and radial metric coefficients. The description of black holes from the thermodynamic point of view will be discussed in Section \ref{sec:4}, and finally Section \ref{sec:5}  is devoted to conclusions.

\section{RSch-AdS mathematical background}
\label{sec:2}

 Noncommutative geometry established based on this fact that Quantum Gravity has an uncertainty principle which prevent one from measuring position to accuracy better than that given by Plank length. This property of geometry removes the infinities that appear and cause bad cut-off behavior of field theory of gravity due to quantum fluctuation in the vicinity of origin and its main role is to replaces point like structures with a smeared objects \cite{key-2,key-11}. The commutator can be written in the form as bellow:

\begin{equation}\label{equ1}
 [X^\mu ,X^\nu]=i\theta^{\mu\nu},
\end{equation} 

where $\theta^{\mu\nu}$ is an antisymmetric tensor that acts as fundamental cell discretization of spacetime just similar to what Planck constant $\hbar$ roles in phase space. The approach to noncommutative quantum field theory and a procedure known as coherent state formalism, modifies the structure of Feynman propagator curing the short distance behavior of point like structure as well as curing divergence that appear under various forms in General Relativity\cite{key-3,key-12}. Therefore a black hole that is defined based on noncommutativity geometry in order to resolving singularity of the origin in Anti de Sitter coordinates is called Regular Schwarzschild Anti-de-Sitter (RSch-AdS) black hole mentioned in the last part of introduction section in which its negative cosmological constant is written as: $\Lambda=-\frac{3}{b^2}$, $b$ is the curvature radius of AdS space, and a matter source or mass density in that vacuum space has been replaced with a\emph{ Gaussian  distribution of matter} in the form of

\begin{equation}\label{equ2}
	\rho(r)=\frac{M}{(4\pi\theta)^{\frac{3}{2}}}\exp^{-(\frac{r^2}{4\theta})},
\end{equation}

where $M$ is \emph{diffused} throughout a region with a linear and minimal lenth $\sqrt{\theta}$ instead of being perfectly localized in a point due to uncertainty encoded in the coordinate spacetime commutator\cite{key-9,key-10,key-11}.
 For next purposes, we rescale the variables as folllows:
     
  \begin{equation}\label{equ3}   
    \tilde{r}=\frac{r}{\sqrt{\theta}},\quad \tilde{M}=\frac{M}{\theta},\quad \tilde{\Lambda}=\theta\Lambda,\quad \tilde{G}=\sqrt{\theta}G,\quad \rho(\tilde{r})=\sqrt{\theta}\rho(r), 
  \end{equation}  
   
  then the mass density can be rewritten as:

\begin{equation}\label{equ4}
  \rho(\tilde{r})=\frac{\tilde{M}}{(4\pi)^\frac{3}{2}}\exp^{-(\frac{\tilde{r}^2}{4})}
\end{equation}

\begin{figure}[h]
	\centering
	\includegraphics[width=0.6\textwidth]{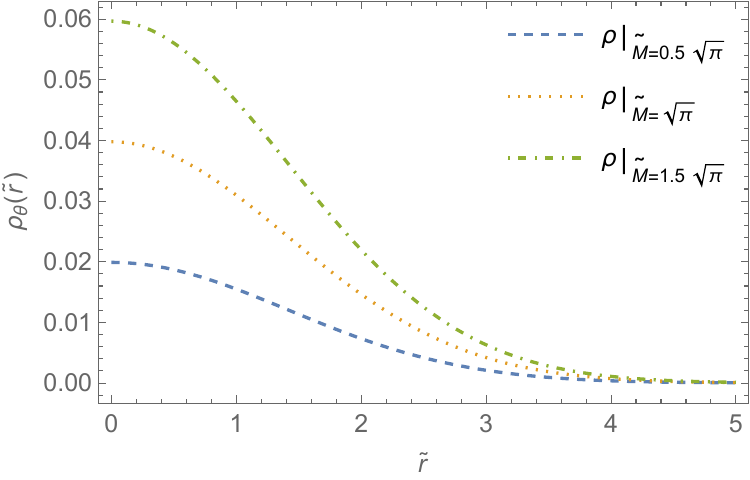}
	\caption{The mass density profile, Equ. \ref{equ4}, decreases exponentially from its maximum value for different values of mass so that it becomes almost zero for $\tilde{r}> 4$. }\label{Fig1}
\end{figure}

This type of matter distribution simulates the noncommutativity of spacetime through the parameter $\theta$, for explanation a natural ultraviolet space-time cut-off (see Fig. \ref{Fig1}).

The covariant conservation condition of energy-momentum tensor requires that $T^{\mu\nu}_{\quad;\nu}=0$, therefore we have

\begin{equation}\label{equ5}
	T^{0}_{\quad 0 }=T^{r}_{\quad r}=-\rho(\theta), \qquad T^{\vartheta}_{\quad \vartheta}=-\rho(r)-\frac{r}{2}\partial_{r}\rho(r).
\end{equation}

The spherically symmetric solution of Einstein's equation with this energy-momentum tensor and the cosmological constant $\Lambda$ is given by line element \cite{key-13}:

\begin{equation}\label{equ6}
	ds^2=-f(r)dt^{2}+\frac{dr^{2}}{f(r)}+r^{2}d\Omega^2,
\end{equation}

where

\begin{equation}\label{equ7}
	f(r)=1+\frac{r^{2}}{b^{2}}-\frac{\omega M}{r}\gamma\left(\frac{3}{2},\frac{r^{2}}{4\theta}\right),
\end{equation}

and $d\Omega^{2}=d\vartheta^{2}+\sin^{2}\vartheta d\varphi^{2}$. The role of cosmological constant is played by $b$ through $\frac{1}{b^2}=-\frac{\Lambda}{3}$, $\omega=2G_{\text{N}}/\Gamma(\frac{3}{2})$, in which $G_{\text{N}}$ is four-dimensional Newton's constant and $\gamma(\frac{3}{2},\frac {r^{2}}{4\theta})$, the lower incomplete Gamma function, is defined generally in the form of $\gamma(n,x)=\int_{0}^{x}t^{n-1}e^{-t}dt$. For the rest of this work we choose a particular geometric where units are defined as $\sqrt{\theta} G_{\text{N}}=\hbar=c=1$, for an asymptotically AdS-BH  case in four dimensions \cite{key-14}. Rewriting the $f(r)$ according to Equ. \ref{equ3} leads to:

\begin{equation}\label{equ8}
  g_{tt}(\tilde{r})=1-\frac{\tilde{\Lambda}}{3}\tilde{r}^{2}-\frac{2\tilde{M}}{\tilde{r}\Gamma(3/2)}\gamma(\frac{3}{2},\frac{\tilde{r}^{2}}{4}).
\end{equation}

 Profile of $g_{tt}(\tilde{r})$ for different values of $\tilde{M}$ and $\tilde{\Lambda}$ (see Fig. \ref{Fig2}) shows the regularity of spacetime for RSch-AdS metric and assures that this spacetime is geodesically complete \cite{key-15}. This metric function, $g_{tt}(\tilde{r})$, for RSch-AdS vs radial distance $\tilde{r}$ shows that for a given mass, for example $\tilde{M}=0.5\sqrt{\pi}$, event horizon of BH depending on the value of cosmological constant occurs at different radii, i.e., for $\tilde{\Lambda}=-0.1, -0.01$  \text{and} -0.001, event horizon located at $\tilde{r}_{\text{H}}\simeq 1.65, 15.18\quad \text{and}\quad \gg 20$ respectively, Fig. \ref{Fig2}-a. For a fixed value of cosmological constant i.e., $\tilde{\Lambda}=-0.01$, and different values of mass i.e., $\tilde{M}=0.5\sqrt{\pi}, \sqrt{\pi}\quad \text{and}\quad 1.5\sqrt{\pi}$, the event horizons are located at $\tilde{r}_{\text{H}}=16.33, 15.5\text{and}\quad 13.5 $ respectively, Fig. \ref{Fig2}-b. This means that for larger values of mass or the cosmological constant, the event horizon radius occurs at smaller distance than when the mass or the cosmological constant have smaller values and the event horizon radius is larger. On the other hand the event horizon of the Sch-AdS metric can be obtained when $\theta\rightarrow 0$ as follows: 
 
 \begin{figure}[]
	\centering
	\subfloat[]{\includegraphics[width=0.45\textwidth]{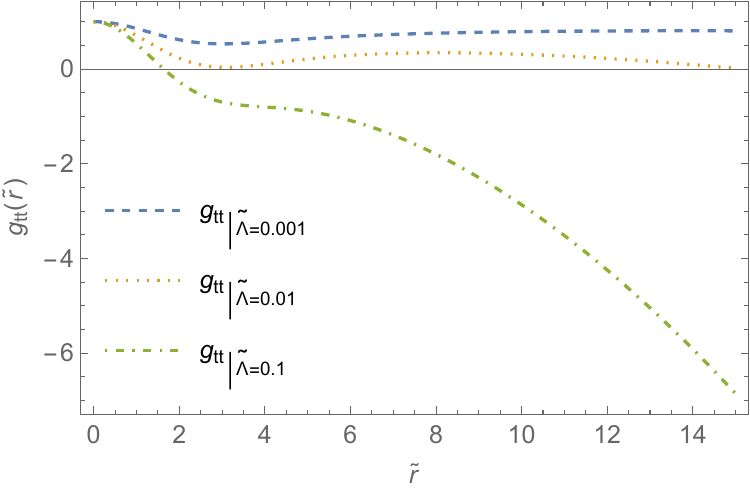}}
	\quad
	\subfloat[]{\includegraphics[width=0.45\textwidth]{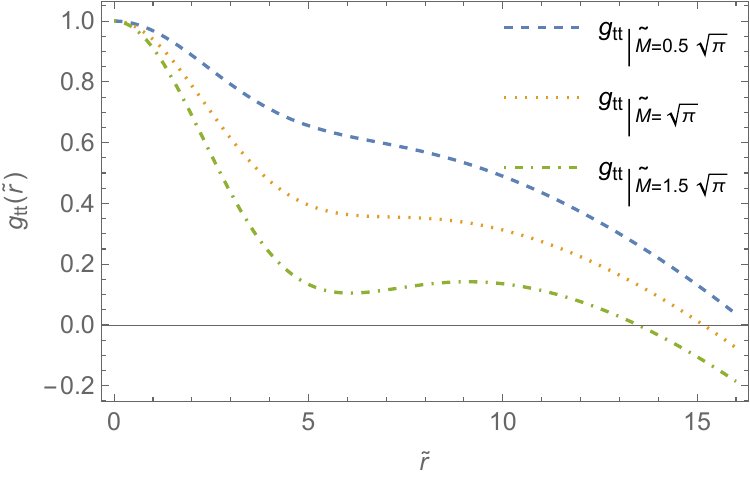}}
	\caption{Time metric function, $g_{tt}(\tilde{r})$, for RSch-AdS spacetime vs radial distance $\tilde{r}$ for $\tilde{M}=0.5\sqrt{\pi}$ and different values of cosmological constant (a) and for $\tilde{\Lambda}=-0.01$, and different values of mass i.e., $\tilde{M}=0.5\sqrt{\pi}, \sqrt{\pi}\quad \text{and}\quad 1.5\sqrt{\pi}$ (b).}\label{Fig2}
 \end{figure}
 
  \begin{figure}[]
	\centering
	\subfloat[]{\includegraphics[width=0.45\textwidth]{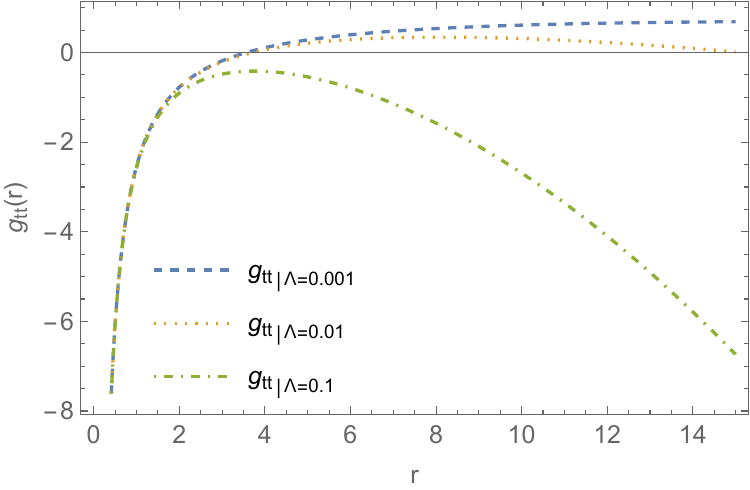}}
	\quad
	\subfloat[]{\includegraphics[width=0.45\textwidth]{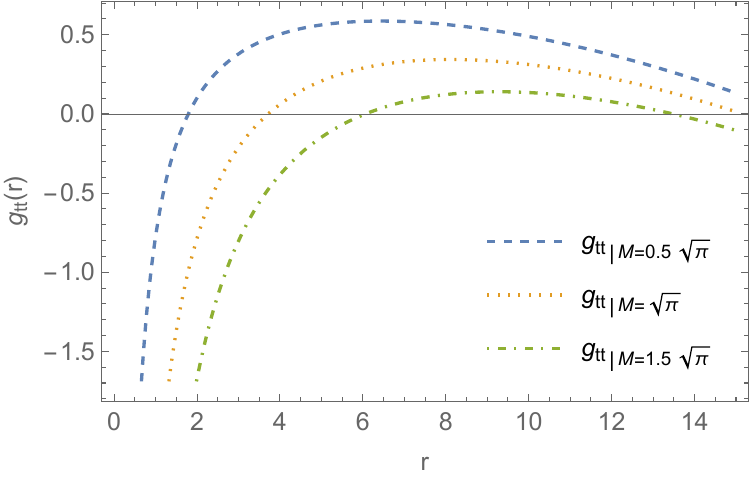}}
	\caption{Time metric function, $g_{tt}$, for Sch-AdS spacetime vs radial distance $r$ for a fixed mass, for example $M=\sqrt{\pi}$, and different values of $\Lambda=-0.001,-0.01\quad \text{and} -0.1$ (a) and for a fixed $\Lambda=-0.01$  and $M=0.5\sqrt{\pi},\sqrt{\pi}\quad\text{and}\quad 1.5\sqrt{\pi}$ (b).}\label{Fig3}
   \end{figure} 
 
   \begin{equation}\label{equ9}
     g_{tt}(r_{0})=1-\frac{\Lambda}{3}r_{0}^{2}-\frac{2M}{r_{0}}. 
   \end{equation}
      
   The graphs related to Equ. \ref{equ9} show that depending on the changes of the values of $\Lambda$ and $M$ as parameters, we can determine the limits imposed on the horizon radius (see Fig. \ref{Fig3}). As we can see when the mass of the BH considered as a fixed value and $\Lambda$ is allowed to have different values, BH has no event horizon for larger values of $\Lambda$ and has an inner event horizon $r_{-}$ as well as an outer event horizon $r_{+}$. On the other hand when the cosmological constant is fixed and mass changes, it is seen that for all mass values, there are two event horizon, one of them is $r_{-}$, known as inner event horizon, and the other, $r_{+}$, is outer event horizon.  
          
   Later in section \ref{sec:4} we will return to Equs. \ref{equ8}and\ref{equ9}, related profiles and their interpretations when we discuss the aspects of BH thermodynamics for RSch-AdS space-time.

\section{Geodesics}
\label{sec:3}
\subsection{Effective potential energy}
\label{subsec:3-1}

Before doing anything about geodesics, it is necessary to have a discussion about energy level, $E$, effective potential energy, $V_{\text{eff}}$, and angular momentum, $L$, related to any massive or massless particle whose its motion is to be considered. First we focus on this fact that for a black hole background and an equatorial plane, $\vartheta=\frac{\pi}{2}$, there are two \emph{conserved quantities} for the Killing fields i.e., $\partial_{\text{t}}$ and $\partial_{\phi}$, the particle energy $E$ and angular momentum $L$ along each geodesics which are obtained through Hamiltonian approach:

\begin{equation}\label{equ10}
    E=p_{t}=-f(r)\dot{t}, \hspace{2cm}L=p_{\phi}=r^{2}\dot{\phi},
  \end{equation}

and second is the effective potential energy, $V_{\text{eff}}$, which for metric mentioned in Equ. \ref{equ7} reads as\cite{key-16, key-17}: 

 \begin{align}\label{equ11}
  V_{\text{eff}}(\tilde{r})=& f(\tilde{r})\left(\frac{L^{2}}{\tilde{r}^{2}}+\varepsilon\right)\nonumber\\
   =& \left(1-\frac{\tilde{\Lambda}}{3}\tilde{r}^{2}-\frac{4\tilde{M}}{\tilde{r}\sqrt{\pi}}\gamma\bigg(\frac{3}{2},\frac{\tilde{r}^{2}}{4}\bigg)\right)\left(\frac{L^{2}}{\tilde{r}^{2}}+\varepsilon\right),
  \end{align}
  where we have put $\theta=\pi$ and $\varepsilon=0$ and $+1$ correspond to massless and massive particles respectively.
  As we can see, effective potential is a function of spacetime metric and also includes parameters related to the motion i.e., cosmological constant, black hole mass and angular momentum, so it is used as a powerful tool to predict different kind of motion for the particle around a heavy mass like a black hole. In the first look at the effective potential plots, we can see several features of particle motion under the influence of this potential. For example, circular motion of the particle in the equatorial plane located in the zero(s) or turning point(s) of this potential. Needless to say that potential plays the same role for motion of the particle as the potential in classical mechanics for one- dimensional\cite{key-18}.

\begin{figure}[]
	\centering
	\subfloat[]{\includegraphics[width=0.5\textwidth]{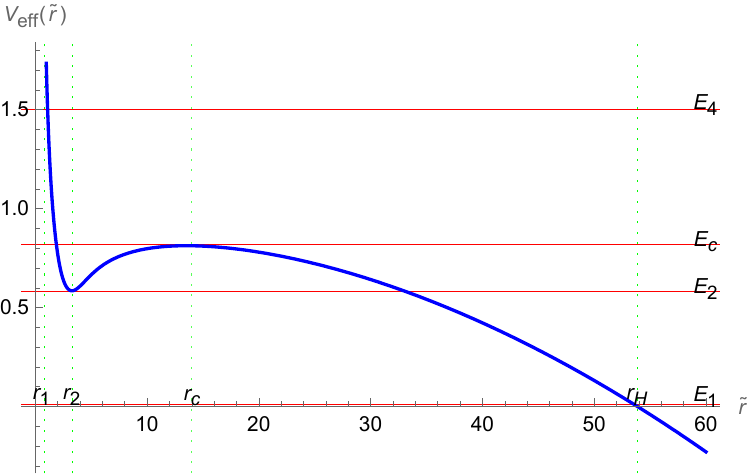}}
    \subfloat[]{\includegraphics[width=0.5\textwidth]{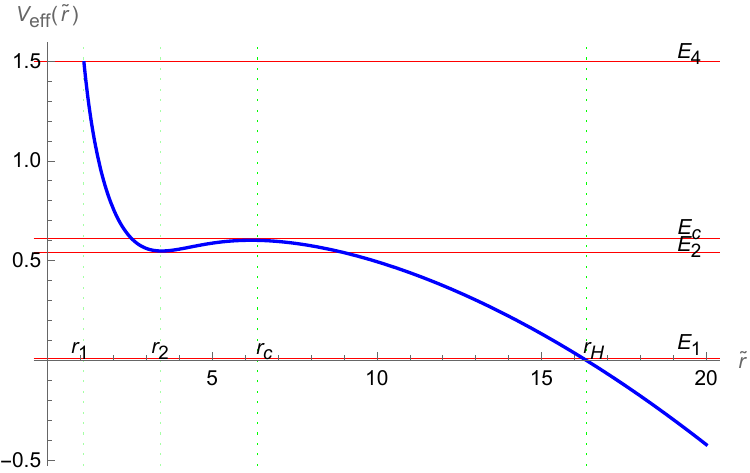}}
    \quad
    \subfloat[]{\includegraphics[width=0.5\textwidth]{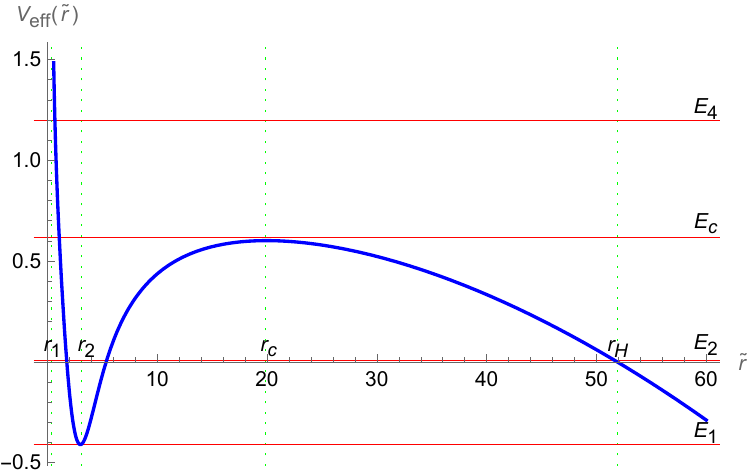}}
    \subfloat[]{\includegraphics[width=0.5\textwidth]{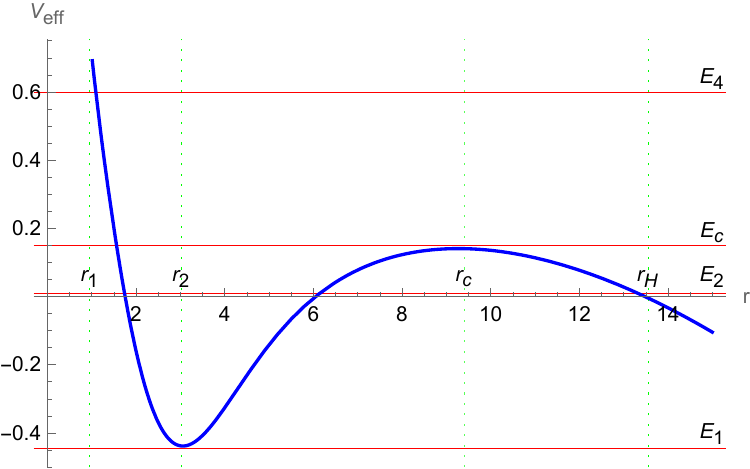}}
	\caption{Effective potential vs $\tilde{r}$ for $L=1$, $\tilde{M}=0.5\sqrt{\pi}$, $\tilde{\Lambda}=-0.001$ (a) and previous values but $\tilde{\Lambda}=-0.01$ (b). Effective potential vs $\tilde{r}$ for $L=0.5$, $\tilde{M}=1.5\sqrt{\pi}$, $\tilde{\Lambda}=-0.001$ (c) and previous values but $\tilde{\Lambda}=-0.01$ (d).}\label{Fig4}
   \end{figure}

   Here, let us have some interpretation about the type of motion for the particle and determine point(s) or region(s) in which particle can have stable or unstable orbits in more details. By having a quick glance at the effective potential plots in general, for example Fig. \ref{Fig4}-a, one can deduce that depending on the particle energy level, $E$, ($E^{2}=\dot{r}^{2}+V_{\text{eff}}$), and for zero radial velocity, $\dot{r}_{0}=0$, the following motions for the particle are possible:
  
  \begin{itemize}
  
    \item For $E>0$ the particle falls into the event horizon.
    \item For $E=0$ and $r=r_{\text{H}}$ particle's path is a closed circle and its orbit is stable. 
    \item For $E<0$ and $r>r_{\text{H}}$, outside the event horizon, the motion of the particle is precessional and its orbit unstable.
   
  \end{itemize}
  
  \subsection{Geodesics equations for RSch-AdS spacetime}
  \label{subsec:3-2}

Now we are in a position that can determine the geodesics of real particles around a black hole for the metric given by Equ. \ref{equ7}.
As we know geodesics is the generalization of path motion from a straight line in Newtonian mechanics to a \emph{straight one in a curved spacetime} in relativity which minimizes the distance between two points. In other words, a freely moving or falling particle always moves along a geodesics in relativity. The geodesics equations of motion in a 4-D curved spacetime for a given world line described by $(x^{\eta}=x^{\eta}(\tau), \eta=0,1,2,3)$ is:

\begin{equation}\label{equ12}
	\frac{d^{2}x^{\eta}}{d\tau^{2}}+\Gamma^{\eta}_{\mu\nu}\frac{dx^{\mu}}{d\tau}\frac{dx^{\nu}}{d\tau}=0,
\end{equation}

where $\Gamma^{\eta}_{\mu\nu}=\frac{1}{2}g^{\eta\lambda}\left\{-g_{\mu\nu,\lambda}+g_{\nu\lambda,\mu}+g_{\lambda\mu,\nu}\right\}$ is known as \emph{affine connection} and Greek indices are used for spacetime components, here $x^{\eta}=(t,r,\vartheta, \phi)$.

For a special case in which $\vartheta=\frac{\pi}{2}$, and RSch-AdS spacetime metrics described by Equs. \ref{equ6} and \ref{equ7}, it is easy to show that geodesics equations are:

\begin{equation}\label{equ13}
  \ddot{t}=-\frac{1}{f(\tilde{r})}\left(\frac{d}{d\tilde{r}}f(\tilde{r}) \right)\dot{\tilde{r}}\dot{t}
\end{equation}

\begin{equation}\label{equ14}
	\ddot{\tilde{r}}=\frac{1}{2}f(\tilde{r})\left(\frac{d}{d\tilde{r}}f(\tilde{r})\right)\dot{t}^{2}-\frac{1}{2}f(\tilde{r})\left(\frac{d}{d\tilde{r}}\frac{1}{f(\tilde{r})}\right)\dot{\tilde{r}}^{2}+\tilde{r}f(\tilde{r})\dot{\phi}^{2},
\end{equation}

\begin{equation}\label{equ15}
	\ddot{\phi}^{2}
	=-\frac{2}{\tilde{r}}\dot{\tilde{r}}\dot{\phi},
\end{equation}

where dot is derivative with respect to $\tau$, the proper time. These equations are a system of three second-ordered coupled equations that is difficult to solve directly and should be solved numerically.

\subsection{initial conditions}
\label{subsec:3-3}

Since geodesics paths of real particles described by Equs. \ref{equ13}, \ref{equ14} and\ref{equ15} are a system of three nonlinear equations and of second-order with respect to proper time, so to achieve the solutions of this system of equations, it is necessary to use numerical solution techniques\cite{key-17,key-19,key-20,key-21} and considering effective potential mentioned in Equ. \ref{equ11}. Geodesics equations aforementioned above are parameterized based on proper time,$\tau$, so initial conditions are imposed on $t$, $\tilde{r}$ and $\phi, (\dot{\phi}=L/m\tilde{r}^{2})$. This means that initial values of time coordinate, $t$, radial position, $\tilde{r}$, angular momentum, $L=m\tilde{r}^{2}\dot{\phi}=\tilde{r}^{2}\dot{\phi}$, and their first order derivatives $\dot{t}$, $\dot{\tilde{r}}$ and $\dot{\phi}$ respectively for $\tau=0$ are required. So for this purpose, the required initial conditions are classified as follow:

\begin{figure}[]
	\centering
	\subfloat[]{\includegraphics[width=0.45\textwidth]{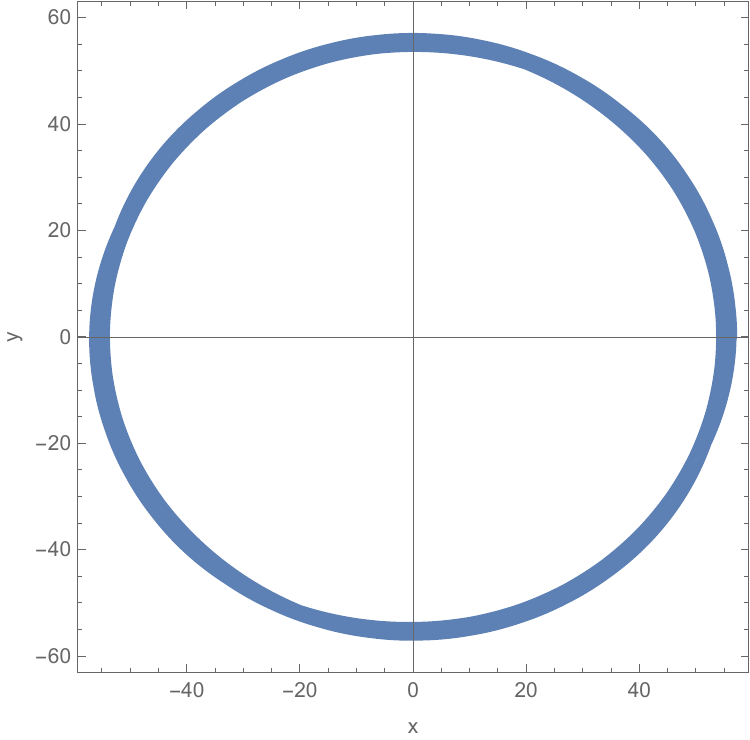}}
	\quad
	\subfloat[]{\includegraphics[width=0.45\textwidth]{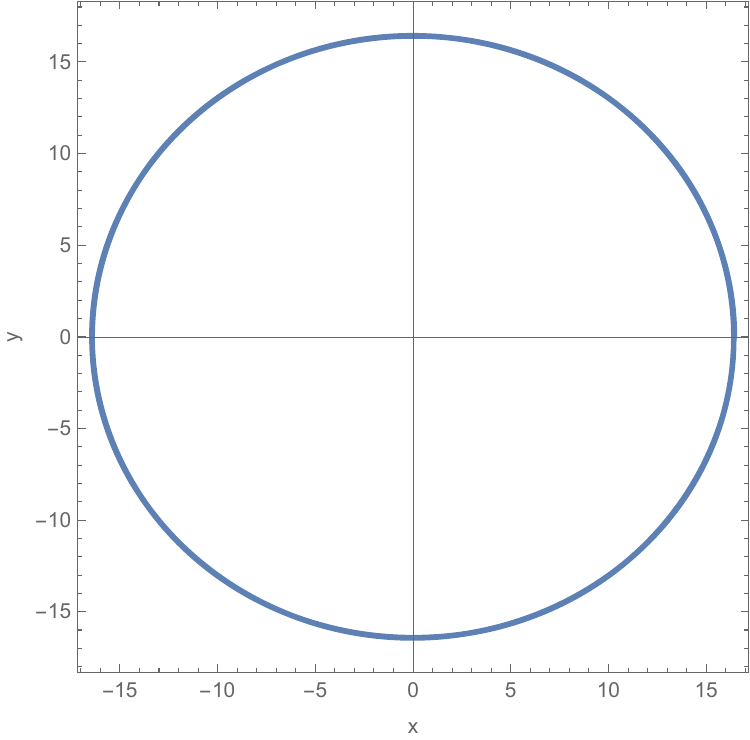}}
	\caption{Particle trajectory plotted for $\tilde{M}=0.5\sqrt{\pi}$, $L=1$, $\tilde{\Lambda}=-0.001$, $\dot{\tilde{r}}_{0}(\tau)=0$, $\tilde{r}_{0}(\tau)=55$  and $\tau=[0,300000]$ as initial conditions (a). Reiteration the previous process but this time for $\tilde{\Lambda}=-0.01$ and $\tilde{r}_{0}(0)=17$ and $\tau=[0,150000]$.}\label{Fig5}
\end{figure}

\begin{figure}[]
	\centering
	\subfloat[]{\includegraphics[width=0.45\textwidth]{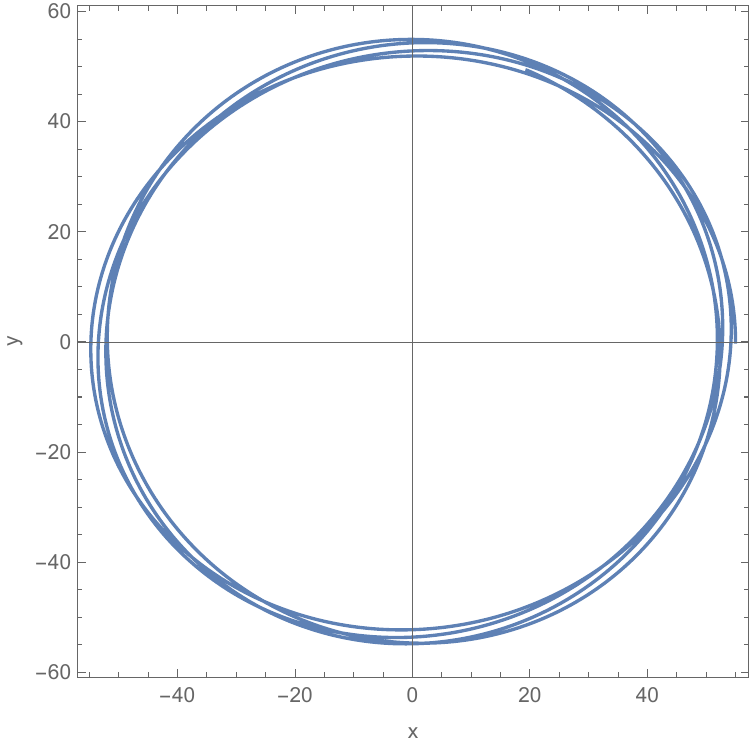}}
	\quad
	\subfloat[]{\includegraphics[width=0.45\textwidth]{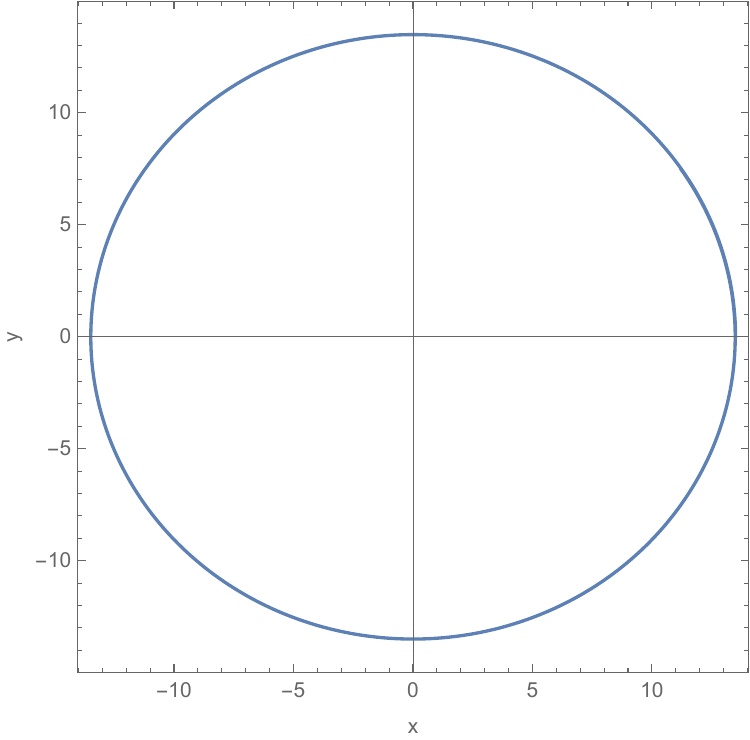}}
	\caption{Particle trajectory plotted for $\tilde{M}=1.5\sqrt{\pi}$, $L=0.5$, $\tilde{\Lambda}=-0.001$, $\dot{\tilde{r}}_{0}(\tau)=0$, $\tilde{r}_{0}(\tau)=54$ and $\tau=[0,300000]$ as initial conditions (a). Reiteration the previous process, but this time for $\tilde{\Lambda}=-0.01$, $\tilde{r}_{0}=13.5$ and $\tau=[0,5000]$.}\label{Fig6}
\end{figure}

\begin{itemize}

	\item $L=0$ resulted in $\phi=\phi_{0}=0$, which means that a particle has a free fall on a straight line from its initial position $(\tilde{r}_{0},\phi_{0})$.
	\item Depending on $L>0$ or $L<0$ and an increasing amount for $\phi$, the particle undergoes counter clockwise and clockwise direction respectively.

	\item For a relatively large amount of angular momentum, for example $L=1$, and choosing the other variables as follows: $\tilde{\Lambda} =-0.001$, $\tilde{M}=0.5\sqrt{\pi}$, for an initial position, $\tilde{r}_{0}(0)=57>r_{\text{H}}=53.8$, and velocity, $\dot{\tilde{r}}_{0}(0)=0$, when proper time changes in interval $0$ and $300000$ shows that the real particle has a precession motion in an orbital with a radial magnitude that varies between $\tilde{r}_{\text{min}}\simeq 53.8$ and $\tilde{r}_{\text{max}}\simeq57.2$ (see Fig. \ref{Fig5}-a). For previous initial condition but this time $\tilde{\Lambda}=-0.01$ and $\tilde{r}_{0}(0)=16.5= r_{\text{H}}$  resulted in a completely different situation in which $V_{\text{eff}}(r=16.5)=0$ and we have a free particle that its orbit approaches to a closing circle of $\tilde{r}=16.5$ (Fig. \ref{Fig5}-b).

	\item For further assurance, we repeat the previous process by choosing new values for the parameters as below: $L=0.5$, $\tilde{M}=1.5\sqrt{\pi}$, $\tilde{r}_{0}(0)=55 >r_{\text{H}}=51.8$, $\dot{\tilde{r}}_{0}(0)=0$ and $\tilde{\Lambda}=-0.001$ as initial conditions. New conditions lead to similar results, i.e., once again (see Fig. \ref{Fig6}-a) the real particle has a precession motion which its radius located in the range $52<\tilde{r}<55$. For previous step but this time $\tilde{\Lambda}=-0.01$ and $\tilde{r}_{0}(0)=13.5$, where $V_{\text{eff}}(\tilde{r}=13.5)=0$, we have a free particle which its motion approaches to a closing circle of radius $\tilde{r}=13.5=r_{\text{H}}$ (Fig. \ref{Fig6}-b).

\end{itemize}

\newpage

\section{Thermodynamics of RSch-AdS }
\label{sec:4}

\begin{figure}[]
	\centering
	\includegraphics[width=0.6\textwidth]{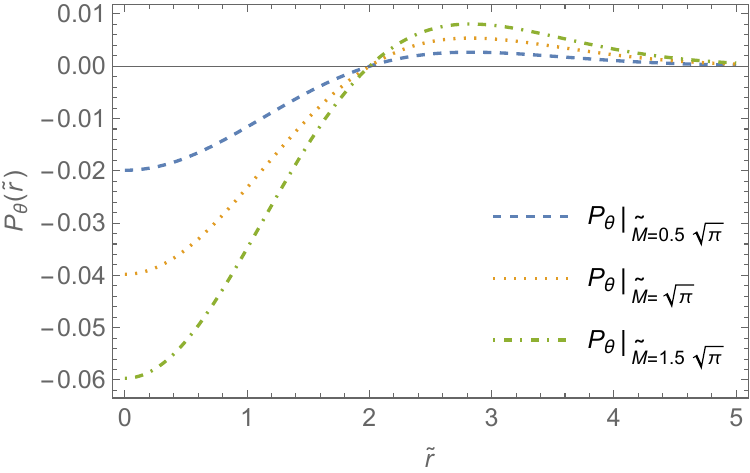}
	\caption{For all attributed values of mass and $\tilde{r}\leq 2.0011$, the tangential pressure is negative and then becomes a little positive but quickly approaches to zero. }\label{Fig7}
\end{figure}

  At present, there are irrefutable evidence for the fact that a black hole can be interpreted as a thermodynamic system in which the role of temperature is played by the surface gravity, i.e., the strength of the gravitational field at the event horizon. Therefore, since thermodynamics provides a description of the system in terms of macroscopic variables without taking into account the microscopic details has been one of the most interesting areas of investigation in recent decades, for instance see \cite{key-8, key-22} and references therein. So study and evaluate some different aspects of RSch-AdS-BH from the point view of thermodynamics lows and compare these features with their corresponding ones in Anti-de Sitter black hole (AdS-BH) will be on the agenda of this section.\\
   First of all, we consider the component $T^{\vartheta}_{\quad \vartheta}=-\rho(\tilde{r})-\frac{\tilde{r}}{2}\partial_{\tilde{r}}\rho(\tilde{r})=p_{\bot}$ mentioned by Equ. \ref{equ5} which is known as the tangential pressure. As one can see from the plot, (see Fig. \ref{Fig7}), for $\tilde{r}\leq2.0011$ and any arbitrary value for $\tilde{M}$, tangential pressure is negative which is defined to be the outward push pressure induced by noncommuting coordinate quantum flactuations \cite{key-2, key-23}. When  $\tilde{r}>2.0011$, for all ranges of mass, pressure becomes slightly positive and then quickly decreases to zero.

As we mentioned before, the line element expressed by Equ. \ref{equ6} can be used to obtain  event horizon radius, ie., $f(r_{H})= 1+\frac{r^{2}_{\text{H}}}{b^{2}}-\dfrac{\omega M}{r_{\text{H}}}\gamma\left(\tfrac32,\tfrac{r^{2}_{\text{H}}}{4\theta}\right)=0$, so it can be read as:

 \begin{equation}\label{equ16}
   \tilde{M}=\frac{1}{2} \tilde{r}_{\text{H}}\left(1-\frac{\tilde{\Lambda}}{3}\tilde{r}^{2}_{\text{H}}\right)\frac{\Gamma(\frac{3}{2})}{\gamma\left(\frac{3}{2},\frac{\tilde{r}^{2}_{\text{H}}}{4}\right)},
 \end{equation}

  where $\gamma({\tilde{r}_{\text{H}}})=\gamma(\frac{3}{2},\frac{\tilde{r}^{2}_{\text{H}}}{4})$ \cite{key-13, key-24}.
  Since $\Gamma[\frac{3}{2}]=\frac{\sqrt{\pi}}{2}=\gamma(\tilde{r}_{\text{H}})+ \Gamma[\frac{3}{2}, \frac{\tilde{r}^{2}_{\text{H}}}{4}]$, then the mass of RSch-AdS-BH in terms of event horizon radius can be written as: 

\begin{equation}\label{equ17}
	M=\frac{1}{2} r_{\text{H}}\left(1-\frac{\Lambda}{3}r_{\text{H}}^{2}\right)+\frac{1}{2} \tilde{r}_{\text{H}}\left(1-\frac{\tilde{\Lambda}}{3}\tilde{r}_{\text{H}}^{2}\right)\frac{\Gamma[\frac{3}{2}, \frac{\tilde{r}_{\text{H}}^{2}}{4}]}{\gamma(\tilde{r}_{\text{H}})}.
\end{equation}

\begin{figure}[h]
	\centering
	\subfloat[]{\includegraphics[width=0.45\textwidth]{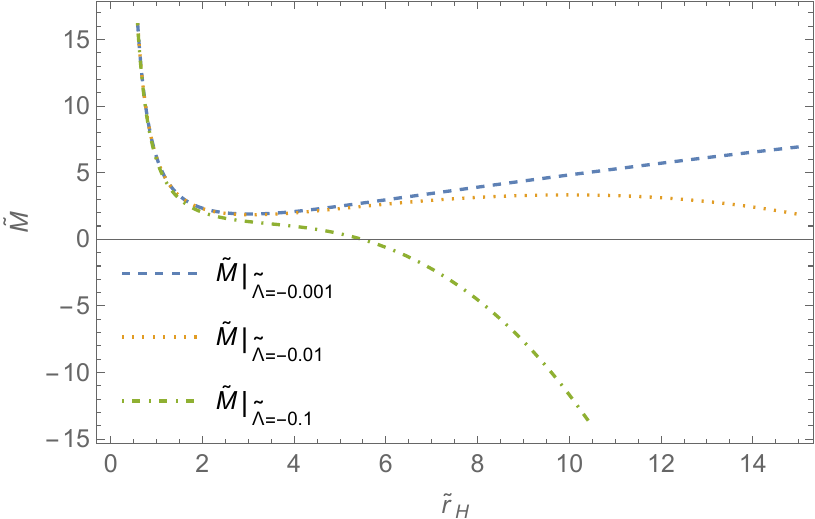}}
	\quad
	\subfloat[]{\includegraphics[width=0.45\textwidth]{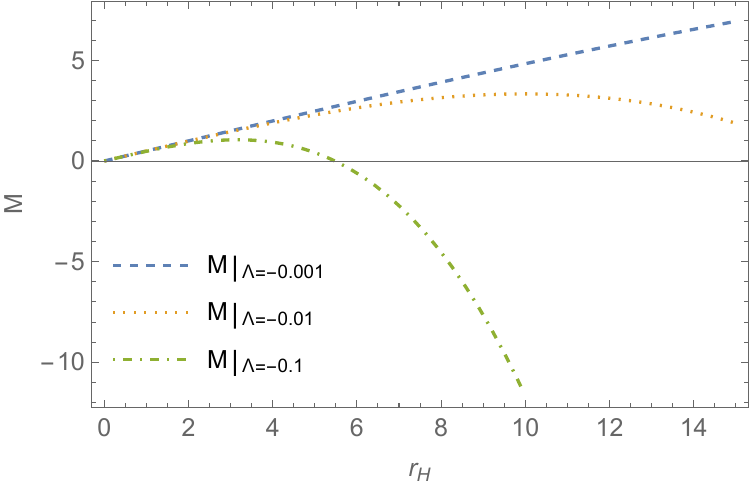}}
	\caption{Mass of RSch-AdS-BH in terms of radius (a) vs mass of AdS-BH in terms of radius (b) for different values of $\Lambda$.}\label{Fig8}
\end{figure}

 The firs term in the right hand of Equ. \ref{equ17}, $\frac{1}{2} r_{\text{H}}(1-\frac{\Lambda}{3}r_{\text{H}}^{2})$, is the mass of BH for AdS spacetime, when $\frac{\tilde{r}^{2}_{\text{H}}}{4}\gg 1$, second term $\frac{1}{2} \tilde{r}_{\text{H}}(1-\frac{\tilde{\Lambda}}{3}\tilde{r}_{\text{H}}^{2})(\frac{\Gamma[\frac{3}{2}, \frac{\tilde{r}_{\text{H}}^{2}}{4}]}{\gamma(\tilde{r}_{\text{H}})})$ brings in $\theta$-correction.
The graph of $\tilde{M}$ as a function of $\tilde{r}_{\text{H}}$ can be used to check the possibility of the existence of event horizon in the sense that for an assigned value of $\tilde{M}$ the horizon(s) are given by the intersections of $\tilde{M}=const$ with the graph $M(\tilde{r}_{\text{H}})$ \cite{key-25}. First in Fig. \ref{Fig8}-a allows us to consider behavior of RSch-AdS-BH mass in terms of radius according to Equ. \ref{equ17}. As one can see the effect of quantum field theory shows that $\tilde{r}_{0}\simeq0.58$ and $\tilde{M}_{0}\simeq  16.19$ are minimum values for black hole formation. $\tilde{M}=1.84$ and $\tilde{r}=2.88$  are considered as minimum values for extremal RSch-AdS-BH for $\tilde{\Lambda}=-0.001$ and $\tilde{\Lambda}=-0.01$. The mass of BH for $\tilde{\Lambda}=-0.1$ and $\tilde{r}_{\text{H}}\geq5.47$ becomes negative which is not acceptable. On the other hand, Fig. \ref{Fig8}-b shows corresponding plot of mass in terms of radius for AdS-BH, which for $\tilde{r}_{\text{H}}\geq 5.47$ these two graphs completely coincide.

 The behavior of mass in terms of entropy for different values of $\Lambda$ is described as $S=\frac{A}{4}$, where $A=4\pi r_{\text{H}}^{2}$ is its surface area, so the mass in terms of entropy for RSch-AdS-BH by considering Equ. \ref{equ16} is as follow:

\begin{equation}\label{equ18}
	\tilde{M}=\frac{1}{2}\sqrt{\frac{\tilde{S}}{\pi}}\left(1-\frac{\tilde{\Lambda}}{3}\frac{\tilde{S}}{\pi}\right)\left\{1+\frac{\Gamma(\frac{3}{2},\frac{\tilde{S}}{4\pi})}{\gamma(\frac{3}{2},\frac{\tilde{S}}{4\pi})}\right\},
\end{equation} 

where this equation reduces to:

\begin{equation}\label{equ19}
 M=\frac{1}{2}\sqrt{\frac{S}{\pi}}\left(1-\frac{\Lambda}{3}\frac{S}{\pi}\right), 
\end{equation}

for AdS-BH.

\begin{figure}[h]
	\centering
	\subfloat[]{\includegraphics[width=0.45\textwidth]{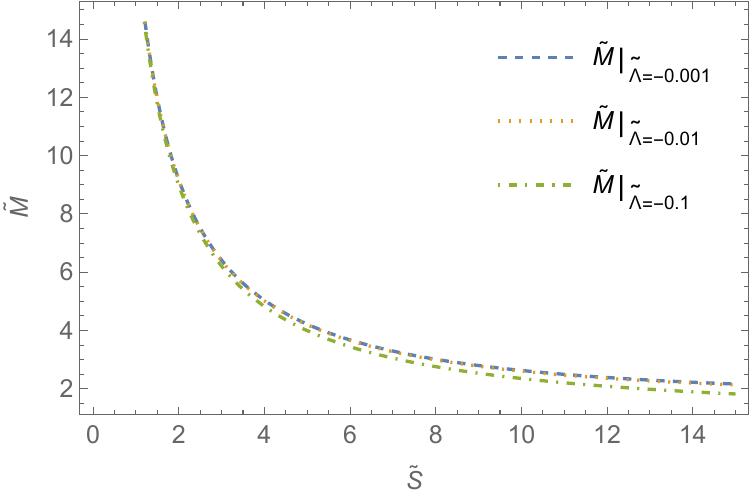}}
	\quad
	\subfloat[]{\includegraphics[width=0.45\textwidth]{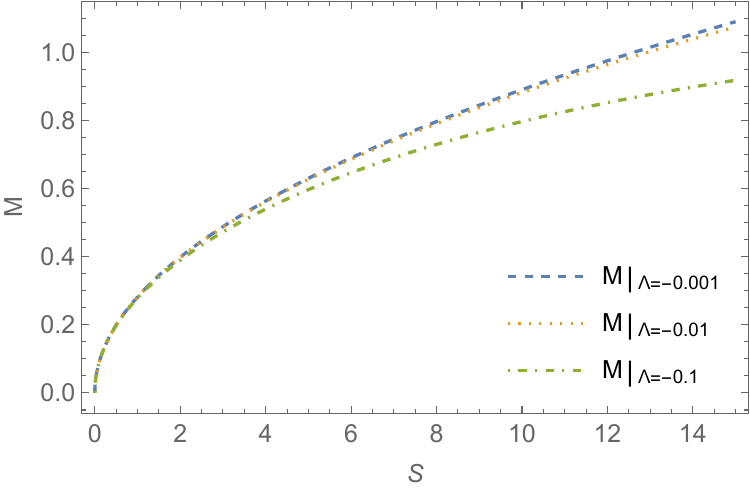}}
	\caption{Mass of RSch-AdS-BH in terms of entropy (a) vs mass of AdS-BH in terms of entropy (b) for different values of $\Lambda$.}\label{Fig9}
\end{figure}

  As we can see from Fig. \ref{Fig9}-a, with increasing the entropy, the mass of RSch-AdS-BH decreases in almost the same way for all three cases of $\tilde{\Lambda}=-0.001, -0.01, -0.1$, so the role of cosmological constant for changes of mass in terms of entropy is negligible but needless to say, with increasing the entropy the mass of RSch-AdS-BH decreases continuously for all aforementioned values of $\Lambda$. In Fig. \ref{Fig9}-b,the profile of mass in terms of entropy for AdS-BH shows a continuous trend of increasing with increasing entropy to coincide with its corresponding in Fig. \ref{Fig9}-a. 
  
The associated temperature to the event horizon of a black hole, as the strength of gravitational field at the event horizon, can be obtained through the relation $T_{\text{H}}=\frac{1}{4\pi} \frac{df(r)}{dr}\mid_{ r=r_{\text{H}}}$. Then we have,

\begin{equation}\label{equ20}
	T_{\text{H}}=\frac{1}{4\pi r_{\text{H}}}\left\{1-\dfrac{\Lambda}{3}r_{\text{H}}^{2}\left(3-r_{\text{H}}\frac{\gamma^{\prime}(r_{\text{H}})}{\gamma(r_{\text{H}})}\right)-r_{\text{H}}\dfrac{\gamma^{\prime}(r_{\text{H}})}{\gamma(r_{\text{H}})}\right\},
\end{equation}

 Where $\frac{d}{dx}\gamma(n,x)=x^{n-1}e^{-x}$ so $\gamma^{\prime}(r_{\text{H}})=\frac{r^{2}_{\text{H}}}{4(\theta)^{3/2}}e^{-r^{2}_{\text{H}}/4\theta}$ \cite{key-24}. By rewriting this relation according to Equ. \ref{equ3}, we have:
 
 \begin{equation}\label{equ21}
   \tilde{T}_{H}=\frac{1}{4\pi \tilde{r}_{H}}\left\{1-\frac{\tilde{\Lambda}}{3}\tilde{r}_{H}^2\left(3-\frac{\tilde{r}_{H}^3}{4}\frac{e^{-\frac{\tilde{r}_{H}^2}{4}}}{\gamma(\frac{3}{2}\frac{\tilde{r}_{H}^2}{4})}\right)-\frac{\tilde{r}_{H}^3}{4}\frac{e^{-\frac{\tilde{r}_{H}^2}{4}}}{\gamma(\frac{3}{2}\frac{\tilde{r}_{H}^2}{4})} \right\},
 \end{equation}
    
    where $\tilde{T}_{H}$ is defined as $\sqrt{\theta}T_{H}$. For a large black hole where $\frac{r^{2}_{\text{H}}}{4\theta}\gg 1 $, this equation reduces to its limit form which is known as \emph{Limit Temperature}(LT) and can be written as:
 
 \begin{equation}\label{equ22}
	T_{\text{H-limit}}=\frac{1}{4\pi r_{\text{H}}}\left(1-\Lambda r^{2}_{\text{H}}\right).
\end{equation}

\begin{figure}[]
	\centering
	\subfloat[]{\includegraphics[width=0.45\textwidth]{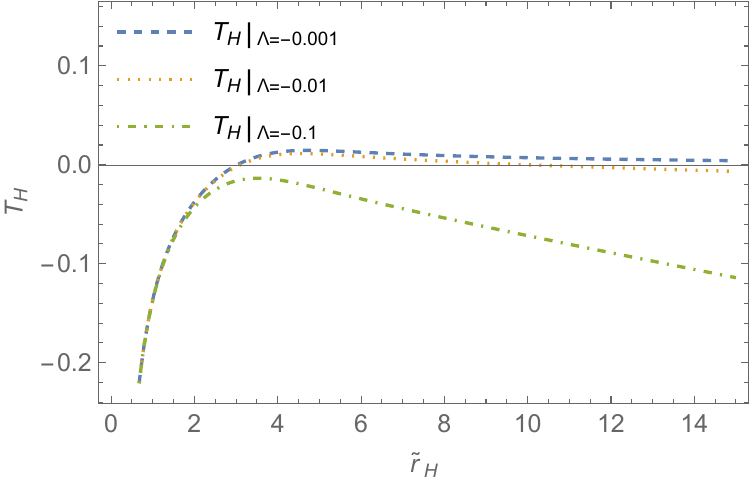}}
    \subfloat[]{\includegraphics[width=0.45\textwidth]{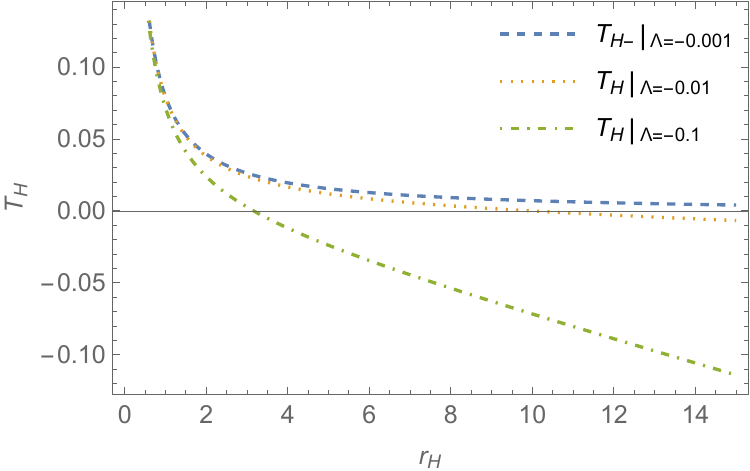}}
	\caption{Black hole event horizon temperature for different values of $\Lambda$ in terms of $r_{\text{H}}$, (a) for RSch-AdS-BH case and (b) for AdS-BH one.}\label{Fig10}
\end{figure}

  Since the event horizon surface temperature of a black hole can not fall below zero, so it can not be justified wherever the chosen values of cosmological constant leads to a negative temperature of the black hole surface and this condition imposes restriction on cosmological constant and black hole radius. Here is the RSch-AdS-BH surface temperature, $\tilde{T}$, vs radius,$\tilde{r}$ , Fig. \ref{Fig10}-a, that can be vary by choosing different values of $\tilde{\Lambda}$ according to bellow: The temperature corresponding to  $\tilde{\Lambda}=-0.1$ is always negative which can not be accepted. $\tilde{r}=3.11$ and $\tilde{r}=9.98$ respectively are considered as minimum and maximum event horizon radius of black hole, in which temperature is positive while $\tilde{\Lambda}=-0.01$. Finally for $\tilde{\Lambda}=-0.001$, in the range of $ 3.038\leq \tilde{r}_{\text{H}}\leq 15$, temperature is  positive with a maximum of $\tilde{T}_{H}=0.0147$. The profile related to the minimum temperature (LT) indicates that the maximum temperature, $T_{H}=0.13$, happens at $r_{\text{min}}=0.602 $ independent of cosmological constant and then we have a rapid decrease in temperature with the increase of radius according to the following: For radii $r_{\text{H}}=3.18$,  $r_{\text{H}}=10.05$ and $r_{\text{H}}>15$ corresponding to $\Lambda=-0.1$, $\Lambda=-0.01$ and $\Lambda=-0.001$ respectively the surface temperature of the BH reaches to zero. (see Fig. \ref{Fig10}-b).

On the other hand, using the entropy is another straightforward way, suing the laws of thermodynamics, to express the temperature of event horizon. According to the first low of thermodynamics, we have:

\begin{equation}\label{equ23}
  dM=Tds+\Phi dQ,
\end{equation}
 where $\Phi$ is electrostatic potential and $Q$ is electric charge. Then for a neutral BH, temperature of event horizon in terms of entropy by considering Equ. \ref{equ18} reads as:
 
 \begin{align}\label{equ24}
  \tilde{T}=& \frac{\partial \tilde{M}}{\partial \tilde{S}}\\ \nonumber
   =& \frac{1}{4\sqrt{\pi \tilde{S}}} \bigg[\left(1-\frac{\tilde{\Lambda}}{\pi}\tilde{S}\right)\left(1+\frac{\Gamma(\frac{3}{2},\frac{\tilde{S}}{4\pi})}{\gamma(\frac{3}{2},\frac{\tilde{S}}{4\pi})}\right) 
   - \frac{1}{8\pi^{\frac{3}{2}}}\tilde{S}^{\frac{3}{2}}\left(1-\frac{\tilde{\Lambda}}{3}\frac{\tilde{S}}{\pi}\right)\frac{e^{-\frac{\tilde{S}}{4\pi}}}{[\gamma(\frac{3}{2},\frac{\tilde{S}}{4\pi})]^{2}}\bigg].
 \end{align}

\begin{figure}[]
	\centering
	\subfloat[]{\includegraphics[width=0.45\textwidth]{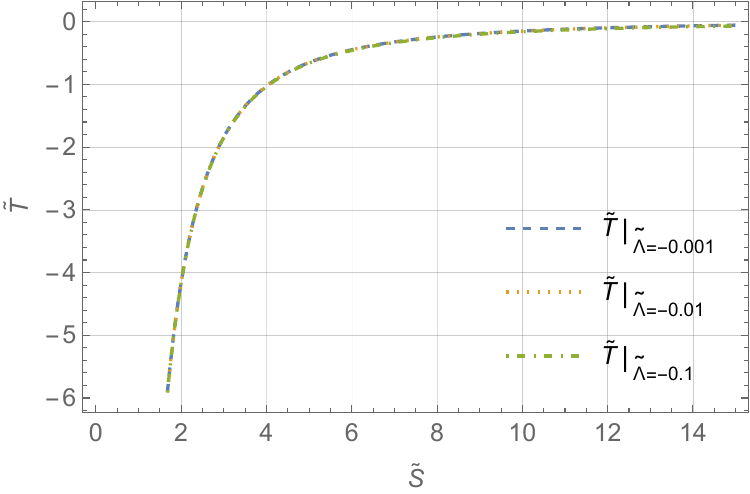}}
    \subfloat[]{\includegraphics[width=0.45\textwidth]{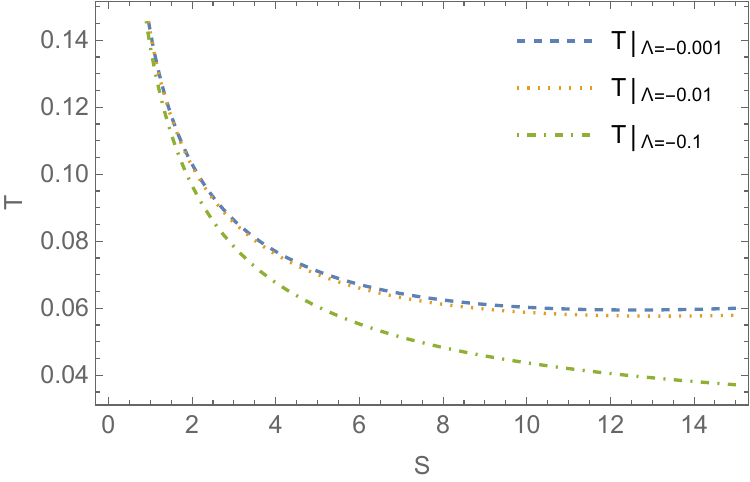}}
	\caption{Black hole event horizon temperature for different values of $\Lambda$ in terms of $S$, for RSch-AdS case (a) and for AdS one (b).}\label{Fig11}
\end{figure}

 The graph of temperature in terms of entropy shown in Fig. \ref{Fig11}-a states that the minimum temperature of Rsch-Ads-BH  is $\tilde{T}_{0}\simeq -5.91$ for $\tilde{S}_{0}\simeq 1.68$. The profile shows that $\tilde{T}$, regardless of the value of $\tilde{\Lambda}$, i.e., $\tilde{\Lambda}=-0.001, -0.01$ and  $-0.1$, increases steeply with the increase of entropy. It can be seen that when $\tilde{S}> 4$, temperature changes uniformly in the range $-5\quad\text{to}\quad 0 $. Equ. (\ref{equ24}) for large values of $S$ reduces to the following form:
  
  \begin{equation}\label{equ25}
  T=\frac{1}{4\sqrt{\pi S}}\left(1-\frac{\Lambda}{\pi}S\right), 
  \end{equation}  
 
 which describes behavior of temperature with changes of entropy for AdS-BH as it is shown in Fig. \ref{Fig11}-b. 
 
In the meantime checking the stability of the black hole is of particular importance, which requires checking its heat capacity, which is related to its surface temperature and entropy through the following formula,

\begin{equation}\label{equ26}
  \tilde{C}=\tilde{T}\frac{\partial \tilde{S}}{\partial \tilde{T}}=\frac{\tilde{T}}{\frac{\partial \tilde{T}}{\partial \tilde{S}}} .
\end{equation}

By considering Equs. \ref{equ24} and \ref{equ26}, the heat capacity equation can be rewritten as follow:

\begin{align}\label{equ27}
  \tilde{C} =&-\frac{1}{4\sqrt{\pi \tilde{S}}} \bigg[\left(1-\frac{\tilde{\Lambda}}{\pi}\tilde{S}\right)\left(1+\frac{\Gamma(\frac{3}{2},\frac{\tilde{S}}{4\pi})}{\gamma(\frac{3}{2},\frac{\tilde{S}}{4\pi})}\right)\nonumber 
   - \frac{1}{8\pi^{\frac{3}{2}}}\tilde{S}^{\frac{3}{2}}\left(1-\frac{\tilde{\Lambda}}{3}\frac{\tilde{S}}{\pi}\right)\frac{e^{-\frac{\tilde{S}}{4\pi}}}{[\gamma(\frac{3}{2},\frac{\tilde{S}}{4\pi})]^{2}}\bigg] \\ \nonumber
   &\times\bigg\{ \frac{1}{16}\bigg[\frac{1}{\sqrt{\tilde{S}}}\left(\frac{1}{\tilde{S}}+\frac{\tilde{\Lambda}}{\pi}\right)\frac{1}{\gamma\left(\frac{3}{2},\frac{\tilde{S}}{4\pi}\right)}+\frac{1}{4\pi^{\frac{3}{2}}}\left(1-\frac{\tilde{\Lambda}}{\pi}\tilde{S}\right)\frac{e^{-\frac{\tilde{S}}{4\pi}}}{\big[\gamma\left(\frac{3}{2},\frac{\tilde{S}}{4\pi}\right)\big]^{2}}\bigg] \\ \nonumber \nonumber
   & +\frac{1}{32\pi^{2}}\bigg[\left(1-(\frac{2}{3}\frac{\tilde{\Lambda}}{\pi}+\frac{1}{4\pi})\tilde{S}+\frac{\tilde{\Lambda}}{12\pi^{2}}\tilde{S}^{2}\right)\frac{e^{-\frac{\tilde{S}}{4\pi}}}{\big[\gamma\left(\frac{3}{2},\frac{\tilde{S}}{4\pi}\right)\big]^{2}}\\ 
  &-\frac{\tilde{S}^{\frac{3}{2}}}{4\pi^{\frac{3}{2}}}\left(1-\frac{\tilde{\Lambda}}{3\pi}\tilde{S}\right)\frac{e^{-\frac{\tilde{S}}{2\pi}}}{\big[\gamma\left(\frac{3}{2},\frac{\tilde{S}}{4\pi}\right)\big]^{3}}\bigg]\bigg\}^{-1}.
\end{align}

\begin{figure}[h]
	\centering
	\subfloat[]{\includegraphics[width=0.45\textwidth]{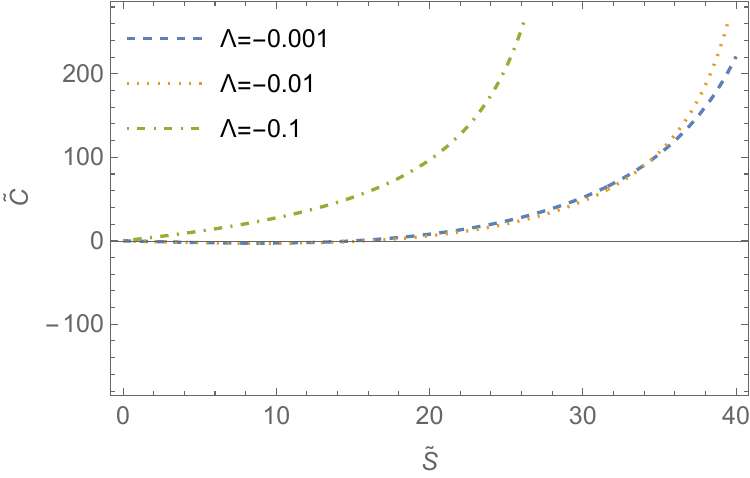}}
	\quad
	\subfloat[]{\includegraphics[width=0.45\textwidth]{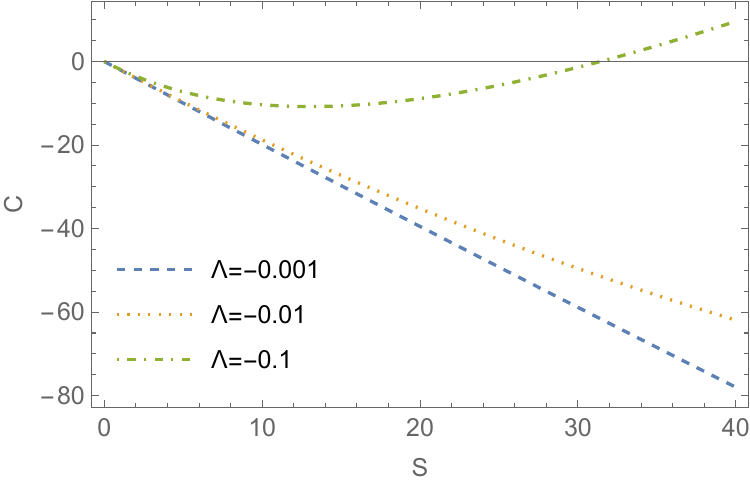}}
	\caption{Heat capacity in terms of entropy for different values of $\Lambda$, for RSch-AdS black hole (a) and AdS-black hole (b) . }\label{Fig12}
\end{figure}

Behavior of heat capacity for a RSch-AdS-BH in terms of $S$ is shown in Fig. \ref{Fig12}-a for different values of $\Lambda$. As we know, a black hole is in its unstable state where the heat capacity is negative and vice versa it is locally stable where heat capacity is positive, therefore for $\tilde{\Lambda}=-0.1$, heat capacity is positive for all values of entropy and we have always a stable black hole. When cosmological constant is chosen to be $\tilde{\Lambda}=-0.01\quad \text{or} -0.001$ , a phase change occurs at $S>16$ and $S>15$ respectively which means heat capacity becomes positive and black hole is in a stable state. For the case where the condition $S\gg 1$ is satisfied, the heat capacity mentioned above reduced as follows:

\begin{equation}\label{equ28}
  C=-2S\left(\frac{1-\frac{\Lambda}{\pi}S}{1+\frac{\Lambda}{\pi}S}\right),
\end{equation}
 
 where we expect a black hole to have a stable condition only for $\Lambda=-0.1$ and $S\geq 31.5$, as shown in Fig. \ref{Fig12}-b. 

 One can use Gibbs free energy to evaluate the equilibrium of a black hole which is in the form:
 \begin{equation}\label{equ29}
   \tilde{G}=\tilde{M}-\tilde{T}\tilde{S}.
 \end{equation}
  
 Substituting Equs. (\ref{equ18}) and(\ref{equ24}) into the last question gives:
  
  \begin{equation}\label{equ30}
    \tilde{G}=\frac{1}{4}\sqrt{\frac{\tilde{S}}{\pi}}\bigg[\left(1+\frac{\tilde{\Lambda}}{3}\frac{\tilde{S}}{\pi}\right)\left(1+\frac{\Gamma(\frac{3}{2},\frac{S}{4\pi})}{\gamma(\frac{3}{2},\frac{S}{4\pi})}\right)
    +\frac{1}{8\pi^{\frac{3}{2}}}\tilde{S}^{\frac{3}{2}}\left(1-\frac{\tilde{\Lambda}}{3}\frac{\tilde{S}}{\pi}\right)\frac{e^{-\frac{\tilde{S}}{4\pi}}}{[\gamma(\frac{3}{2},\frac{\tilde{S}}{4\pi})]^{2}}\bigg].
  \end{equation}
  
\begin{figure}[h]
	\centering
	\subfloat[]{\includegraphics[width=0.45\textwidth]{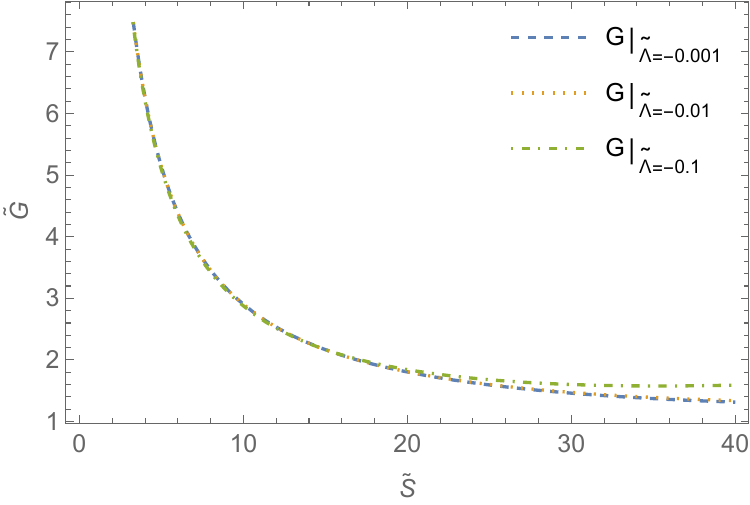}}
	\quad
	\subfloat[]{\includegraphics[width=0.45\textwidth]{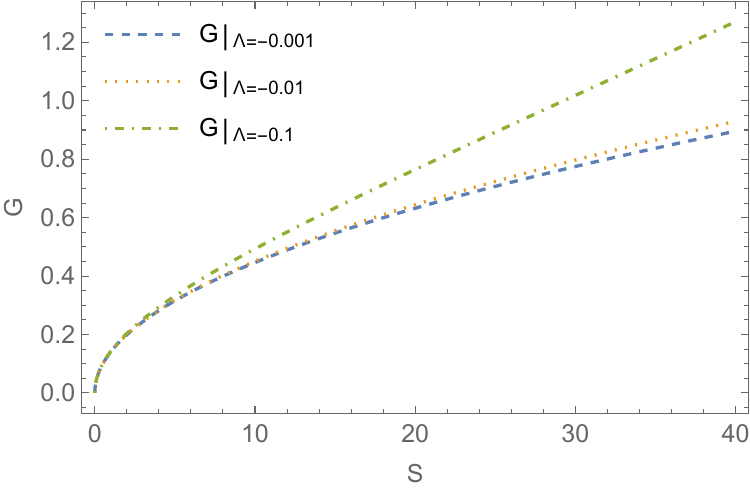}}
	\caption{Gibs free energy in terms of entropy for different values of $\Lambda$, for RSch-AdS black hole (a) and AdS-black hole (b) . }\label{Fig13}
\end{figure}

   Gibbs free energy profile for RSch-AdS block hole shown in Fig. \ref{Fig13}-a, as a function of entropy illustrates that this quantity decreases steeply with increasing entropy when $S\leq 10$. Increasing the entropy i.e.,  $S> 10$ shows that the changes of Gibbs free energy in terms of entropy will be approximately in the range of $1 \leq G \leq 2.5$,  which means we have the best conditions for a black hole equilibrium.     
   Finally for the limit state, that is, when the condition $S\gg 1$ is satisfied, Equ. \ref{equ30} reads as:
   
   \begin{equation}\label{equ31}
    G=\frac{1}{4}\sqrt{\frac{S}{\pi}}\left(1+\frac{\Lambda}{3}\frac{S}{\pi}\right),
   \end{equation}
   
   which describes Gibbs free energy for AdS-BH and its plot shown in Fig. \ref{Fig13}-b tells us that Gibbs free energy changes in terms of entropy is monotonically increasing in a way which coincides with its corresponding i.e., RSch-AdS, in limit state.    

\newpage
\section{Conclusion}
\label{sec:5}

\begin{enumerate}
  \item In Fig. \ref{Fig1} and \ref{Fig7} mass density and tangential pressure described based on Gaussian distribution of matter are plotted in terms of radial distance i.e., rescaled linear size $\frac{r}{\sqrt{\theta}}=\tilde{r}$ for different values of $\tilde{M}$. we can see that for these situations cosmological constant $\Lambda$ has no role. For different values of $\tilde{M}$ negative tangential pressure indicates outward push pressure induced by non-commutative coordinate quantum flactuations.
      
  \item Time metric coefficients, $g_{tt}(\tilde{r})$ for RSch-AdS vs $\tilde{r}$ shown in Fig. \ref{Fig2} implies that when mass is considered as a constant value, for example $\tilde{M}=0.5\sqrt{\pi}$, the event horizon radius is smaller for larger values of cosmological constant and vice versa. On the other hand, for fixed values of cosmological constant, heavier black holes have smaller event horizon radii.   
      
  \item The movement of real particles around the black hole described by RSch-AdS spacetime depending on the choice of appropriate quantities, for example, mass, angular momentum, cosmological constant, initial position and velocity and considering effective potential can be described by closed and stable circles or open and unstable circles with a precession motion (see Figs.\ref{Fig4},\ref{Fig5},\ref{Fig6}).  
      
  \item  Event horizon radius of a BH can be obtained by setting the time metric coefficient equal to zero and then by slightly shifting the expressions, the mass of black hole can be obtained in terms of event horizon radius. The mass of BH as a function of $r_{\text{H}}$ can be used to check the possibility of the existence of $r_{H}$ in the sense that for an assigned value of $M$, the horizon(s) are given by the intersection of $M=const$ with the graph of $M(r_{\text{H}})$. Plot shown in Fig. \ref{Fig8} indicates that the larger the cosmological constant, the smaller the radius of the horizon and vice versa.
         
  \item As shown in Fig. \ref{Fig9}, the decrease the mass of the black hole, regardless of the chosen values of the cosmological constant, leads to an increase in entropy    This issue shows a close connection between entropy as a classical thermodynamic quantity, on one side and  the geometry of spaccetime and mass as relativistic-quantum quantities on the other side. 
           
  \item Surface temperature of RSch-AdS-BH in terms of event horizon radius is unexpectedly negative for smaller values of $\tilde{r}_{\text{H}}$ and different values of $\tilde{\Lambda}$, as the radius increases, the surface temperature becomes slightly positive for some values of $\tilde{\Lambda}$ (see Fig. \ref{Fig10}). On the other hand, according to Fig. \ref{Fig11}, when temperature is expressed as a function of entropy, for all selected values of $\tilde{\Lambda}$, the surface temperature of the black hole is negative.   
        
    \item Heat capacity in terms of entropy shown in Fig. \ref{Fig12} indicates that black hole is unstable when $\tilde{\Lambda}=-0.001$ or $\tilde{\Lambda}=-0.01$ and $\tilde{S}\leq 15.4$ due to the negative heat capacity and for other cases black hole is in a stable situation. 
                     
  \item Profile of Gibbs free energy in terms of entropy shows that for different values of cosmological constant and $\tilde{S}>20$ is suitable condition for black hole to be in equilibrium (see Fig. \ref{Fig13}).
      
  \item Finally, from the results obtained in this work, it can be seen as a general conclusion that the investigation of the thermodynamic conditions of the black hole based on quantum considerations in the AdS spacetime for a Gaussian distribution of mass leads to better results compared to when only the relativistic conditions are considered alone.     
\end{enumerate}


%
%



\end{document}